# Measurement of the Rate Capability of Resistive Plate Chambers

Burak Bilki[d], John Butler[b], Ed May[a], Georgios Mavromanolakis[c,1], Edwin Norbeck[d], José Repond[a], David Underwood[a], Lei Xia[a], Qingmin Zhang[a,2]

[a]*Argonne National Laboratory, 9700 S. Cass Avenue, Argonne, IL 60439, U.S.A.*
[b]*Boston University, 590 Commonwealth Avenue, Boston, MA 02215, U.S.A.*
[c]*Fermilab, P.O. Box 500, Batavia, IL 60510-0500, U.S.A.*
[d]*University of Iowa, Iowa City, IA 52242-1479, U.S.A.*

**Abstract.** This paper reports on detailed measurements of the performance of Resistive Plate Chambers in a proton beam with variable intensity. Short term effects, such as dead time, are studied using consecutive events. On larger time scales, for various beam intensities the chamber's efficiency is studied as a function of time within a spill of particles. The correlation between the efficiency of chambers placed in the same beam provides an indication of the lateral size of the observed effects. The measurements are compared to the predictions of a simple model based on the assumption that the resistive plates behave as pure resistors.

## INTRODUCTION

Resistive Plate Chambers (RPCs) are known to be limited in their capability to operate when exposed to high particle intensities. Previous measurements have established the loss in efficiency for the detection of minimum ionizing particles (MIPs) while being irradiated with radioactive sources [1] or exposed to continuous electron beams at varying intensities [2].

This paper explores the time dependence of the performance of RPCs, when exposed to a proton beam of varying intensity. The short-term inefficiency (dead time after a particle crosses the chamber) and the decrease in overall detection efficiency for MIPs as a function of time have been studied in detail. These effects were investigated exploiting the 3.5 second spill structure of the Fermilab test beam. In addition, utilizing the fine granularity of the readout of the chambers, the spatial dependence of the observed loss in efficiency has been studied.

---

[1] Also affiliated with University of Cambridge, Cavendish Laboratory, Cambridge CB3 OHE, U.K.
[2] Also affiliated with Institute of High Energy Physics, Chinese Academy of Sciences, Beijing 100049, China and Graduate University of the Chinese Academy of Sciences, Beijing 10049, China.

This research was performed within the framework of the CALICE collaboration [3], which develops imaging calorimetry for the application of Particle Flow Algorithms (PFAs) [4] to the measurement of hadronic jets at a future lepton collider.

## DESCRIPTION OF THE RESISTIVE PLATE CHAMBERS

The RPCs measured 20 x 20 cm$^2$ and contained two glass plates. The thickness of the glass plates was 1.1 mm and the gas gap was maintained by fishing lines with a diameter of 1.2 mm. Figure 1 shows a schematic of the chamber design.

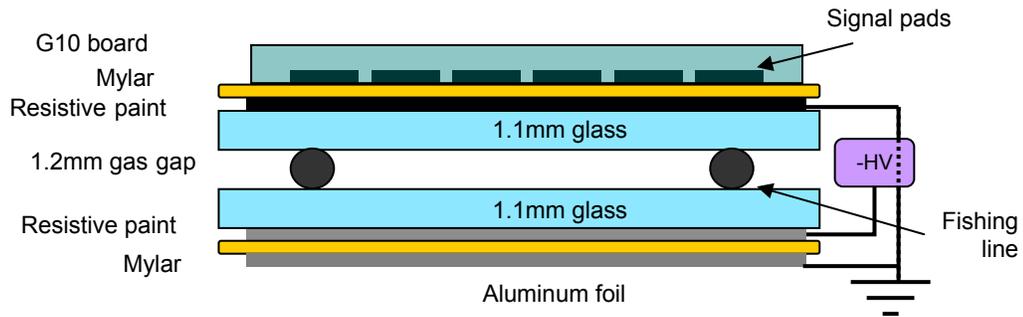

**Figure 1.** Schematic of the chamber design utilized in the present measurements. Not to scale.

The bulk resistivity of the glass was measured to be approximately $\rho \sim 4.7 \cdot 10^{12}$ $\Omega$cm at room temperature. After being heated by the front-end electronics the resistivity was observed to drop to $\rho \sim 3.13 \cdot 10^{12}$ $\Omega$cm. A conductive acrylic paint providing a finite surface resistivity was applied to the outer surfaces of the glass plates. The resistivity of the conductive paint was approximately 1 M$\Omega$/□ and was measured to be uniform to within a factor of two across the surface of the glass plates.

The chambers were operated in saturated avalanche mode with a high voltage setting of 6.3 kV. The gas consisted of a mixture of three components: R134A (94.5%), isobutane (5.0%) and sulfur-hexafluoride (0.5%). Under these operating conditions the fraction of streamers was negligible. For more details on the design and performance of the chambers, see references [5,6].

The electronic readout system was optimized for the readout of large numbers of channels. Each chamber was read out with 16 x 16 readout pads, each with an area of 1 x 1 cm$^2$. In order to avoid an unnecessary complexity of the electronic readout system, the charge resolution of individual pads was reduced to a single bit (digital readout). Event data consisted of a time stamp (with a resolution of 100 ns) and a hit pattern. For more details on the readout system see reference [7].

# TEST BEAM SETUP AND DATA COLLECTION

The chambers were inserted into a hanging file structure. For mechanical stability the chambers were mounted on PVC frames with a hole cut out coinciding with the sensitive area of the chambers. The gap between the frames was 13.4 mm, where 8.3 mm were taken by the chambers and their readout boards. The stack was exposed to the 120 GeV/c primary proton beam at the Meson Test Beam Facility (MTBF) of Fermilab [8].

The beam came in spills of approximately 3.5 second length every one minute. During the spill the beam intensity was close to constant, except at very low rates where the intensity was observed to decrease within a spill. For different runs the intensity of the beam was varied between 500 and 11,000 particles per second.

The readout of the stack was triggered by the coincidence of two large scintillator paddles, each with an area of 19 x 19 $cm^2$, located approximately 2.0 and 0.5 meters upstream. In order to avoid problems with overflowing buffers in the data acquisition system, after each trigger a veto of 1 ms was introduced during which no other triggers were accepted. For a subset of the measurements this veto was reduced to 0.3 ms.

Using the imaging capability of the RPCs the beam spot was measured to be about 6 $cm^2$ in area, independent of beam intensity. The trigger counters were significantly larger than the beam spot and provided a direct measurement of the beam intensity.

# CALCULATION OF THE TIME DEPENDENCE

The following simplified model is used to calculate the RPC response while exposed to a high rate of charged particles. The RPC is assumed to be of infinite lateral size compared to its thickness, so that edge effects can be ignored. Turning on at time t=0, the charged particle flux is assumed to be uniformly distributed over the whole RPC with a rate of f $Hz/cm^2$. The RPC itself consists of three layers: two identical glass plates (thickness $d_2$, voltage drop $u_2$ in the direction of $d_2$, dielectric constant $\varepsilon_2$, and bulk resistivity $\rho$) and a uniform gas gap (thickness $d_1$, voltage drop $u_1$ in the direction of $d_1$, dielectric constant $\varepsilon_1$) between them, as shown in Fig. 2.

The top surface of the RPC is connected to ground and the bottom surface is connected to negative high voltage, at a potential –u, where we assume that the resistance of the power supply is negligible compared to the resistivity of the glass. The inner and outer surfaces of the top (bottom) glass layer have surface charge densities of $\sigma_1$ $C/cm^2$ and $\sigma_2$ $C/cm^2$ ( -$\sigma_1$ $C/cm^2$ and -$\sigma_2$ $C/cm^2$), respectively.

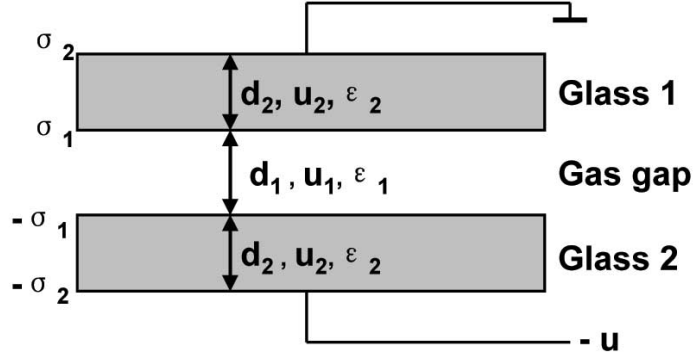

**Figure 2.** Schematic of the chambers including the definition of the electrical constants.

At time t ≤ 0, the signal rate in the gas gap is 0 and the gas gap is not conductive. As a result, the calculation can be treated as an electrostatics problem. Assuming the glass behaves as a pure resistor leads to the following equations, where $\varepsilon_0$ is the vacuum permittivity:

$$u_1 = u, \ u_2 = 0, \ \sigma_1 = \frac{\varepsilon_1 \varepsilon_0 u}{d_1}, \ \sigma_2 = 0, \text{ for } t \leq 0 \tag{1}$$

When a charged particle passes through the gas gap, the initial ionizations will be amplified and an avalanche is formed which deposits charge on the inner surfaces of the glass layers. The average charge deposited depends on the electrical potential applied across the gas gap and shows a threshold behavior. Below turn-on voltage $u_0$, the charge is negligible; above but close to the turn-on voltage $u_0$, it increases linearly with the gap voltage, as shown in the following formula, where c is a constant depending on the gap size and the specific gas being used:

$$q = 0, \text{ for } u_1 \leq u_0 \text{ and } q = c(u_1 - u_0), \text{ for } u_1 > u_0 \tag{2}$$

With a uniform charged particle exposure, a current will flow through the RPC system and modify the surface charge density on all surfaces. The voltages $u_1$ and $u_2$ can be determined from the charge densities in the following way:

$$u_1 = \frac{d_1(\sigma_1 + \sigma_2)}{\varepsilon_1 \varepsilon_0}, \ u_2 = \frac{d_2 \sigma_2}{\varepsilon_2 \varepsilon_0}, \tag{3}$$

where all the voltage drops sum up to be equal to the applied voltage u:

$$u_1 + 2u_2 = u \tag{4}$$

The voltage $u_2$ will introduce a current through the glass plate with current density i (A/cm$^2$), to dissipate the charge deposition from the charged particle signals. The current density i and the voltage $u_2$ are related via Ohm's Law:

$$i = \frac{u_2}{\rho d_2} \tag{5}$$

The charge density on the inner surfaces of the glass layers is modified by the charge deposition from avalanche signals, as well as the current through the glass layers. From charge conservation follows:

$$\frac{d\sigma_1}{dt} = i - fq \tag{6}$$

The solution for $u_1$, the voltage applied on the gas gap for $t \geq 0$, is obtained by combining equations (2) – (6) and using equation (1) as a boundary condition at $t = 0$:

$$u_1 = \frac{2cf\rho d_2(u-u_0)}{1+2cf\rho d_2}e^{-\frac{t}{\tau}} + \frac{u+2cf\rho d_2 u_0}{1+2cf\rho d_2} \quad \text{with} \quad \tau = \frac{\rho\varepsilon_0(2d_2\varepsilon_1 + d_1\varepsilon_2)}{d_1(1+2cf\rho d_2)} \tag{7}$$

From eq. (7) we can see that after the charged particle flux is turned on at $t = 0$, the gap voltage experiences an exponential drop with a time constant $\tau$. After some time $t \gg \tau$ the gap voltage $u_1$ will settle at a constant value, $u_{1f}$, where

$$u_{1f} = \frac{u+2cf\rho d_2 u_0}{1+2cf\rho d_2} \quad \text{and} \quad u > u_{1f} > u_0. \tag{8}$$

The gap voltage is not directly measurable in our experiment. However, the efficiency depends approximately linearly on the operating voltage (= gap voltage at low rate), for efficiencies between 20% and 90% [5]. As a result, the efficiency will experience an exponential drop with the same time constant $\tau$, after turning on the charged particle flux, and will eventually ($t \gg \tau$) settle at a constant value which corresponds to the gap voltage $u_{1f}$.

Quantitative predictions were obtained using the following values for the various parameters in the calculations:

$\varepsilon_0 = 8.85 \times 10^{-12}$ F/m   $\varepsilon_1$ (gas) = 1   $\varepsilon_2$ (glass) = 6
$d_1 = d_2 = 0.115$ cm
$\rho = 3.13 \times 10^{12}$ Ω/cm
$c = 4$ fC/V
$u_0 = 5200$V.

The relation between applied voltage and MIP detection efficiency was taken from measurements.

# MEASUREMENT OF SHORT TERM EFFECTS

Short term effects, i.e. decreases in efficiency after a particle crosses the chambers, were investigated using a run where the data acquisition veto was reduced to 0.3 ms.

In this run the beam intensity was about 670 Hz/cm$^2$ and the average efficiency of the chamber was approximately 55% (see below). The study included only the first chamber in the stack, but the conclusions reached from utilizing the other chambers in the stack were identical.

Figure 3(left) shows the distribution of the time difference between consecutive events, where the times of the individual events have been reconstructed using the time stamps in the event data. The histogram corresponds to pairs of events where the second event has at least one hit in the first chamber, whereas the data points correspond to pair of events where the second event has no hits in the first chamber. In both cases the first event of a pair was required to have at least one hit. The two distributions are seen to be similar in shape.

The efficiency, defined as the number of pairs of events with at least one hit in the second event over the total number of pairs, is shown in Fig. 3 (right) as a function of the time between consecutive events. A fit of the data to a first degree polynomial, shown as a red line in the figure, yields a slope which is consistent with zero. The data show no evidence of a larger inefficiency at shorter compared to longer time differences. This might be due in part to the fact that the beam spot was about 6 cm$^2$ in area and that short term inefficiencies are only expected to occur within lateral distances which are comparable to the gas gap thickness.

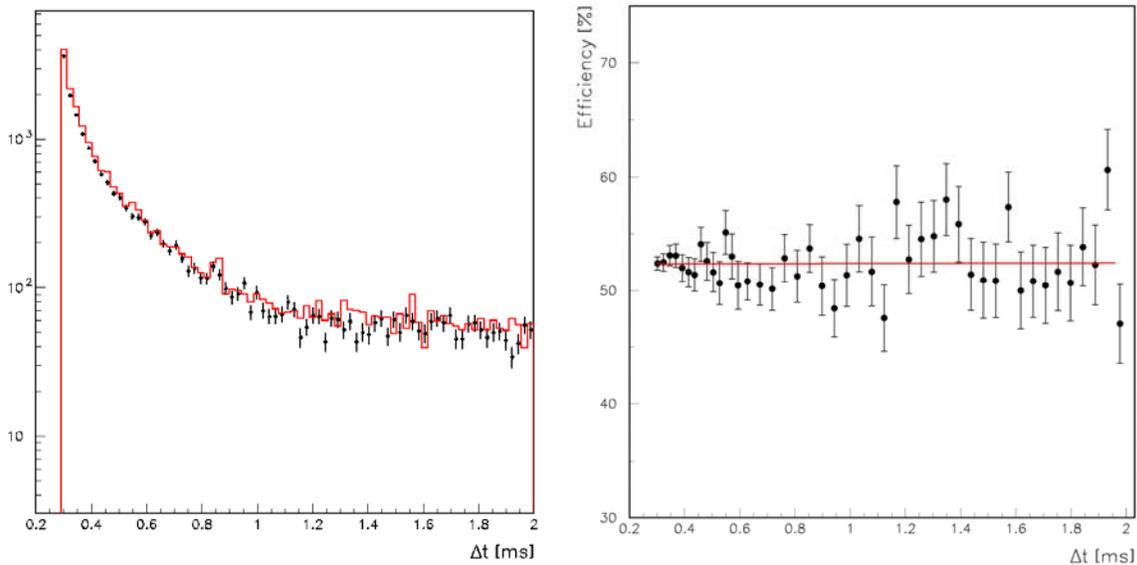

**Figure 3**. Left: Distribution of the time difference between consecutive events. The red histogram (data points) corresponds to pair of events where the second event has at least one recorded hit (no hits) in the first chamber. At least one hit in the first chamber was required in the first of the pair of events. Right: Efficiency as a function of the time difference between consecutive events. The red line represents a fit to a first degree polynomial.

# MEASUREMENT OF THE EFFICIENCY AS A FUNCTION OF BEAM INTENSITY

The efficiency for the detection of MIPs traversing the RPCs has been measured [6] previously using broadband muons, obtained from the primary 120 GeV proton beam together with a 3 m long beam blocker. The beam intensity for these measurements was very low, of the order of 20 – 50 Hz, and so was not expected to affect the performance of the RPCs. In these measurements the efficiency of a given chamber was determined using track segments reconstructed from the hits in the chambers upstream and downstream of the one being tested.

For the present measurement, we define the MIP detection efficiency as the ratio of the number of events with at least one hit in a given chamber to the number of triggered events. This definition of the efficiency has the advantage of being insensitive to correlations in the efficiency possibly present among chambers. On the other hand, due to the loose trigger given by the coincidence of the two scintillator paddles, the efficiency is in general underestimated by including e.g. events where the particle misses the chambers in the stack. Since this study investigates the effect of varying beam intensities on the performance of the chambers, the absolute value of the MIP detection efficiency is only of secondary importance.

To avoid a bias in the calculation of the efficiency from interacting protons, only the first chamber in the stack was used in this study.

The rate of the beam was reconstructed using the number of triggers per spill and was corrected for the size of the beam spot to quote the beam intensity in $Hz/cm^2$, as is customary in the literature. The beam spot was reconstructed using the RPCs themselves and was subject to uncertainties related to the Gaussian shape of the beam and the pad multiplicity of around 1.5 for single MIPs [6]. A systematic error of ±30% was therefore assigned to the rate calculations.

Using the time stamp of the event data, the time elapsed from the beginning of a spill, here named spill time, could be reconstructed. Figure 4 shows the efficiency of the first chamber in the stack as function of spill time for various beam intensities. A clear decrease in efficiency is observed as a function of spill time. As expected from the analytical calculations presented in this paper, the decrease is adequately reproduced by the sum of an exponential and a constant, shown as lines in the figure.

Figure 5 shows the efficiency measured at the beginning and the end of the spills for various beam intensities. The efficiency at the beginning of the spill is around 85%, independent of beam intensity, except at very high beam intensities, where the efficiency decreases already within the first spill-time bin. At large rates the efficiency levels out around 20%.

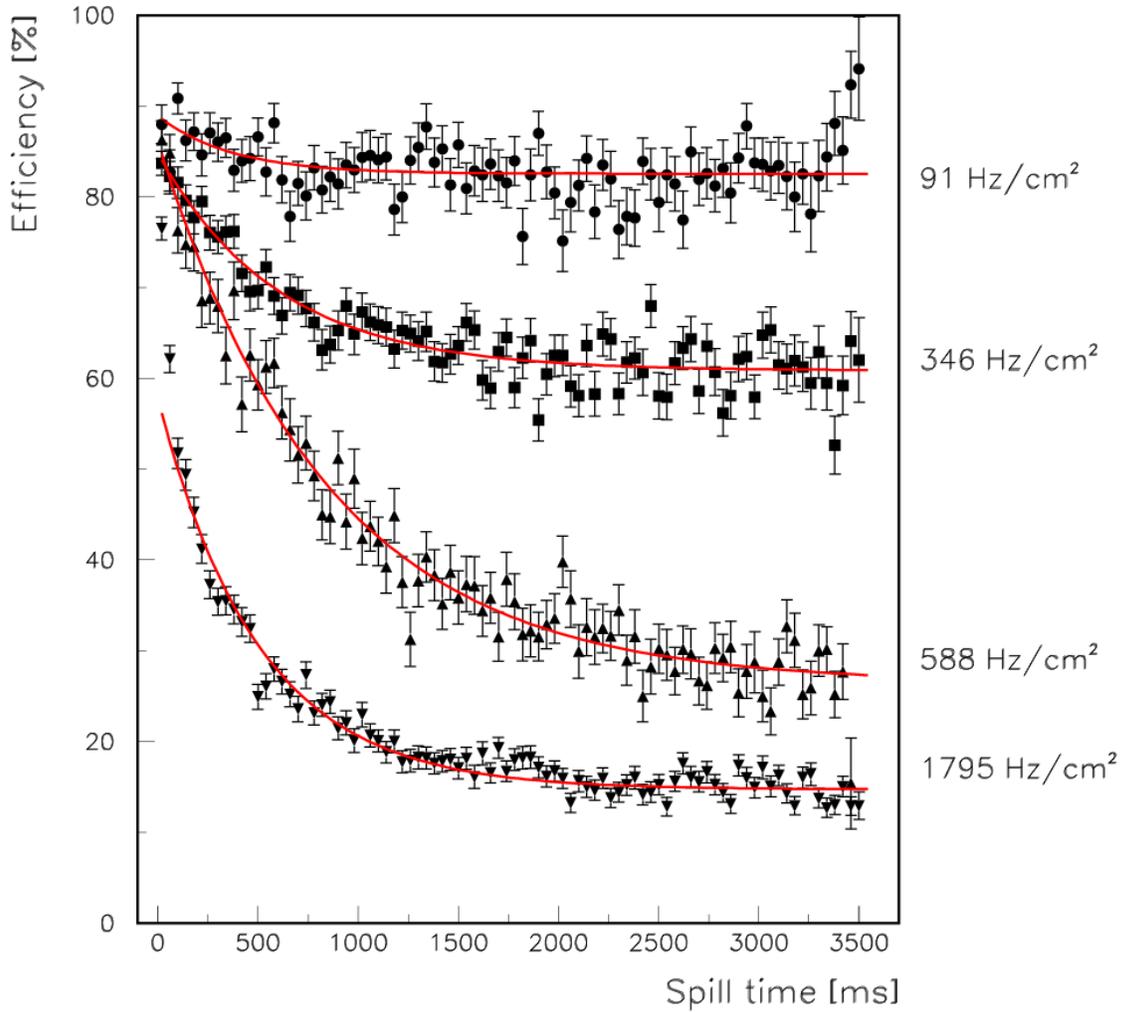

**Figure 4**. MIP detection efficiency as a function of spill time for various beam intensities. The red curves are fits to the data using the sum of an exponential and a constant.

The data in Fig. 5 are compared with the predictions from the analytical calculation. With an assumed turn-on voltage of $u_0$ = 5200 V the general trend of the data is adequately reproduced. However, it needs to be noted, that with a digital readout system the turn-on voltage can not be measured precisely and that the quantitative predictions depend strongly on its value.

The time constant of the exponential in the fit to the data in Fig. 4 is shown in Fig. 6 as a function of beam intensity. Since the loss of efficiency is very small at low beam intensities, the measurements below 300 Hz/cm$^2$ have large uncertainties. Overlaid are the results from the analytical calculations, which adequately describe the data at large beam intensities.

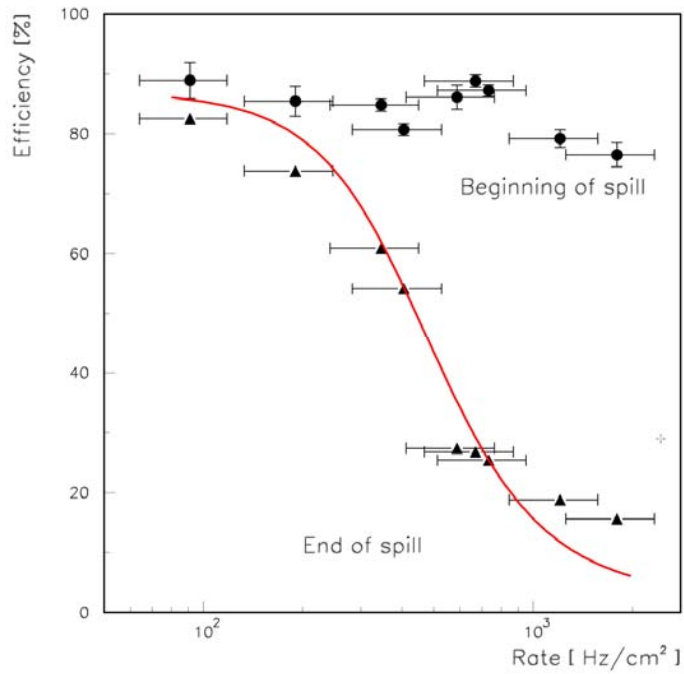

**Figure 5**. MIP detection efficiency as a function of beam intensity at the beginning (squares) and the end (triangles) of a spill. The red line shows the predictions for the constant term of the efficiency.

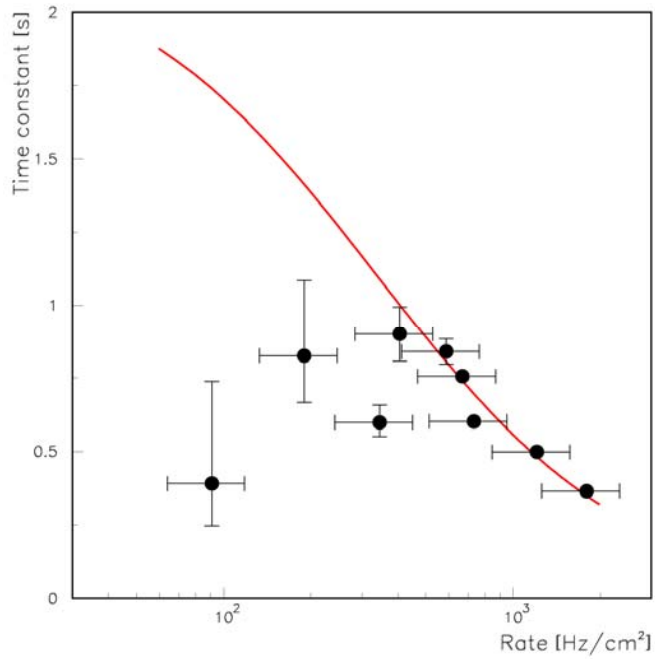

**Figure 6.** Time constant of the exponential decrease in efficiency as a function of beam intensity. The horizontal error bars reflect the uncertainty in the size of the beam spot. The red line represents the predicted values based on the analytical calculations presented in this paper.

# CORRELATION BETWEEN CHAMBERS

Using the first two chambers in the stack, correlations in the inefficiencies of the RPCs were investigated. At the higher beam intensities a strong correlation was observed. Using the data obtained with a beam intensity of ~1800 Hz/cm$^2$, Figure 7a(b) shows the reconstructed position in the second chamber for the case when no (at least one) hit was recorded in the first chamber. In comparison with Fig. 7b, the no-hit data of Fig. 7a appears more focused, suggesting a non-uniform efficiency over the surface of the chamber.

Under the condition that there is at least one hit in the second chamber, the MIP detection efficiency of the first chamber as function of position (reconstructed using the second chamber) is shown in Fig. 7c. The data indicate a clear decrease of efficiency of the chamber at the beam spot compared to the outer regions of the chamber. From this observation we deduce that the rate effects in the RPCs are local to the area of high beam intensities and do not affect the entire chamber surface uniformly. The apparent low efficiency on the edge of the chamber is an artifact of the low statistics in these regions.

# CONCLUSIONS

A stack of four RPCs were exposed to a 120 GeV proton beam of varying intensity at the Fermilab MTBF facility. The data were utilized to study the time dependence of the efficiency for the detection of minimum ionizing particles. The following observations were made:

- Based on the study of consecutive events, no short term inefficiencies with time constants in excess of 0.3 ms were observed. Due to the large beam spot, this study is only sensitive to non-local effects with lateral distances at least an order of magnitude larger than the gas gap size.
- At beam intensities in excess of 100 Hz/cm$^2$, the efficiency is seen to decrease exponentially with time after the beam turns on, until reaching a constant value.
- The time scale for the exponential decrease is of the order of 0.5 second.
- The constant value of the efficiency, reached after the exponential decrease, depends on the beam intensity and is smaller at high rates.
- Based on the study of the correlation between chambers, the beam induced inefficiencies are seen to be local (in the area of the beam spot), rather than affecting the entire chamber.
- A simple calculation based on the voltage drop in the gas gap due to the current flow through the chamber at high rates reproduces the main features of the observed loss in efficiency as a function of spill time.

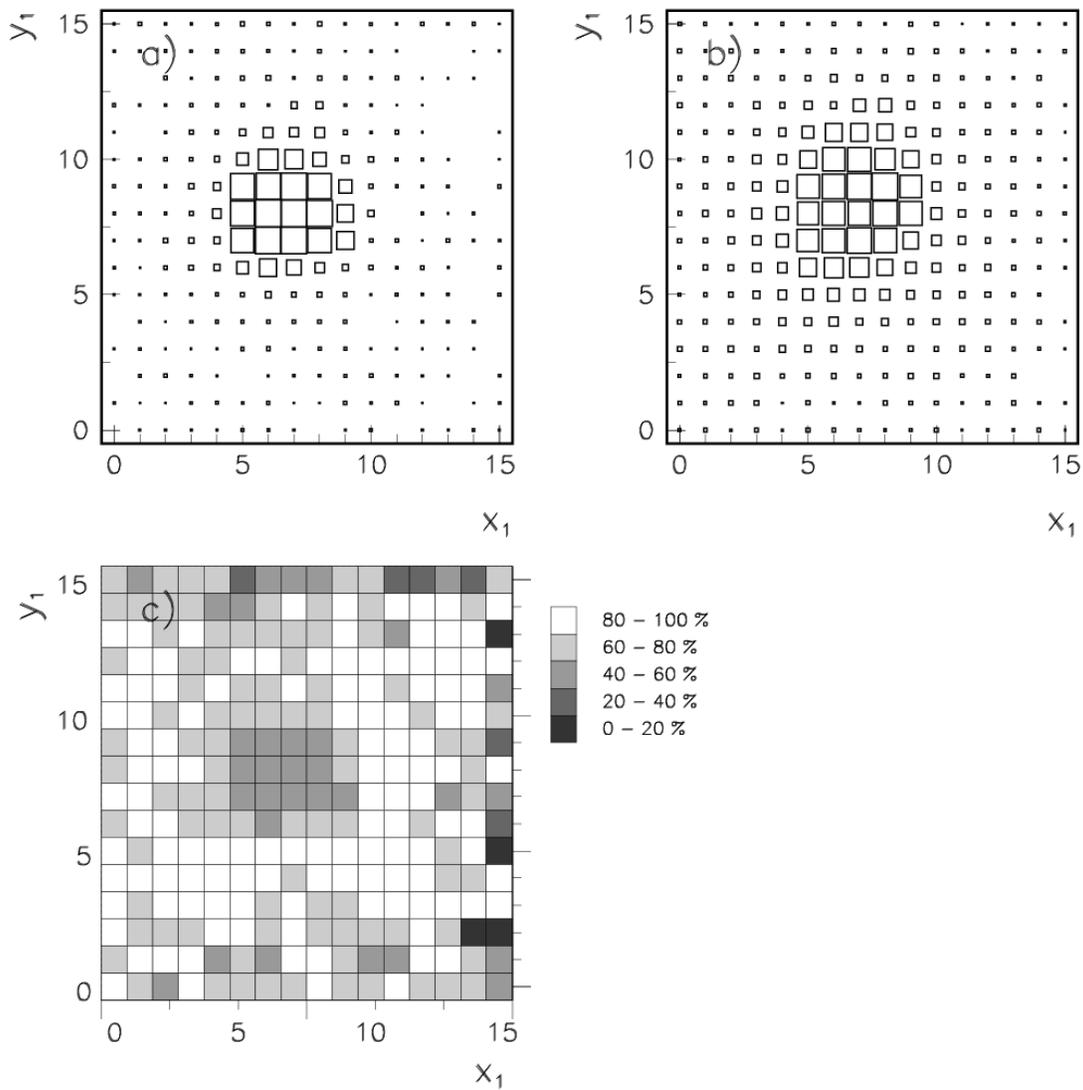

**Figure 7.** Reconstructed position of the proton in the second chamber for a) no hits in the first chamber and b) at least one hit in the first chamber. c) Efficiency of the first chamber when requiring at least one hit in the second chamber. In a given event, the x and y positions are calculated as the centers of gravity of the hits. The uncertainties in the efficiencies are typically better than ±2% at the beam spot and better than ± 8% in the surrounding ring.

## ACKNOWLEDGEMENTS

We would like to thank the Fermilab test beam crew, in particular Erik Ramberg, Doug Jensen, Rick Coleman and Chuck Brown, for providing us with excellent beam. The University of Texas at Arlington ILC group is acknowledged for providing the two trigger scintillator paddles and their associated trigger logic.